\documentstyle[preprint,revtex]{aps}
\math-with-secnums
\begin{document}
\draft
\def\doublespaced{\baselineskip=1.0\normalbaselineskip}
\let\doublespace = \doublespaced
\doublespace
\hfill{UM-P-92/38}

\hfill{OZ-92/10}

\begin{title}
A Non-Commutative Geometric Approach to \\
Left-Right Symmetric Weak Interactions
\end{title}
\author{B. E. Hanlon and G. C. Joshi}
\begin{instit}
Research Centre for High Energy Physics \\
School of Physics \\
University of Melbourne \\
Parkville, Victoria 3052 \\
Australia
\end{instit}
\begin{abstract}
By combining the generalized
exterior algebra of forms over a noncommutative
algebra with the gauging of discrete directions and the
associated Higgs fields, we consider the construction of the bosonic
sector of left-right symmetric models of the form
$SU(2)_{L}\otimes SU(2)_{R}\otimes U(1)$. We see that within
this formalism maximal use can be made of the gauge connection
associated with the non-commutative graded algebra.

\end{abstract}
\newpage
\section{Introduction}
A promising mechanism for the introduction of Higgs fields into
Yang-Mills theory is provided by formally extending space-time to
include additional space-like dimensions. Compactification of these
additional dimensions and the imposition of particular symmetry
properties implied by the internal manifold, such as with the Coset
Space Dimensional Reduction Scheme [1], result in gauge fields
which originally
carried space indices corresponding to the additional dimensions
being realised as scalar fields in four dimensions. The resulting
Higgs potential is in a symmetry breaking form. While this procedure
has an aesthetic appeal, problems associated with the symmetry
breaking scale (i.e. the symmetry breaking is presumed to occur at
the Planck scale) made this mechanism phenomenologically difficult to
reconcile.

Recently, however, an alternate approach has been investigated
by Balakrishna, G\"ursey and Wali [2]. In
this framework the extension of space-time is by finite matrices.
By
defining an exterior algebra of forms over
the noncommutative direct product algebra
of smooth functions on space-time and
hermitian $n\times n$ matrices $\cal A$, a Lagrangian can be
constructed from consideration of the generalized 2-forms. The form of
the Higgs fields is determined by the choice of the exterior
derivative. Furthermore, a
symmetry breaking Higgs potential arises naturally.

The consideration of such an approach was inspired, in part at least,
by new ideas generated by the field of noncommutative geometry
[3] where
there are more than one identical copies of space-time characterized
by a discrete index $p=0,1,2,...$ [4,5].
The simplest example of this is
provided when $p=0,1$, so we have a discrete set of two identical
space-times and the natural appearence of a $Z_{2}$ symmetry. An
$SU(2)\otimes U(1)$ model has been constructed via this approach [5]
which, however, does not make full use of of the inherent symmetries
associated with the gauge connection.

In this paper we seek to combine these new
ideas. By considering the case
of two identical universes with the set of forms described on each by
the generalized algebra $\cal{A}$ we have a way of geometrically
introducing a set of Higgs fields which mimic a set required to break
$SU(2)_{L}\otimes SU(2)_{R}\otimes U(1)$ along the lines of the
standard model. Furthermore, this approach allows for the simultaneous
consideration of a discrete set of extended space-times. We find that
we can make maximal use of the generalized gauge connection and that
the form of the electroweak symmetry breaking Higgs field is
intuitively consistent with
its role as a connection between discrete space-times.
We begin by
reviewing briefly both procedures and then consider a particular model
exhibiting $SU(2)_{L}\otimes SU(2)_{R}\otimes U(1)$ symmetry
which utilizes this approach. We then discuss how the model can be
adjusted so as to incorporate a more realistic symmetry breaking
scheme.
\section{Extended Gauge Theories}
Treating Yang-Mills and Higgs fields at an equal level is an attempt
to understand the symmetry breaking mechanism arising in the standard
model and to constrain some of the parameters associated with it.
Recent work in this field has been done using the ideas of
noncommutative geometry [3].
Within this approach Higgs fields arise from the gauging of discrete
directions. In the case of a $Z_{2}$
symmetry the connection one-form
can be written as a $2\times 2$ matrix [5]
\begin{equation}
w =
\left (
\begin{array}{cc}
\; A \;\;\;\; \phi \;\; \\
\; {\overline{\phi}} \;\;\;\; B \;\;
\end{array}
\right )\;\; .
\label{connect}
\end{equation}

In order to construct field strength tensors and, ultimately,
Lagrangians, a differential operator on the $2\times 2$ matrix
connection (\ref{connect}) must be defined. For a matrix
with differential form coefficients the odd and even parts are resolved
\begin{eqnarray}
X =
\left (
\begin{array}{ll}
\; A \;\;\;\; C \;\; \\
\; E \;\;\;\; B \;\;
\end{array}
\right ) \;\;
\rightarrow
X_{even} =
\left (
\begin{array}{cc}
\; A \;\;\;\; 0 \;\; \\
\; 0 \;\;\;\; B \;\;
\end{array}
\right ) \;\; , \;\;
X_{odd} =
\left (
\begin{array}{rr}
\; 0 \;\;\;\; C \;\; \\
\; E \;\;\;\; 0 \;\;
\end{array}
\right )
\end{eqnarray}
allowing the generalized differential form operator to be defined as
\begin{equation}
dX = i[ \eta_{\gamma} , X_{even}] + i \{ \eta_{\gamma} , X_{odd} \}
+ \left (
\begin{array}{rr}
\;\; dA \;\; -dC \;\; \\
\; -dE \;\;\; dB \;\;
\end{array}
\right ) \;\; .
\end{equation}
This operator, acting on matrices with arbitrary entries, defines a
$Z_{2}$ grading with $\partial X_{odd}=1$ and $\partial X_{even}=0$.
The matrix, $\eta_{\gamma}$, is the most general odd
matrix satisfying $\eta_{\gamma}^{2} = {\bf{1}}$ acting on the space
of $2\times 2$ matrices, which therefore has the form
$ \eta_{\gamma} = cos(\gamma) \tau_{1} + sin(\gamma) \tau_{2} ,
$
where $\tau_{1}, \tau_{2}$ and $\tau_{3}$ are the Pauli matrices. The
operator, $d$, now acts as a derivation:
$ d(X\odot Y) = dX\odot Y + (-1)^{\partial X} X\odot dY , $
where $\partial X$ is the sum of the $Z_{2}$ gradings of $X$ as a
matrix and as a form. The product $\odot$ is defined by
\begin{equation}
 (a\otimes A)\odot (a'\otimes A') = (-1)^{\partial A\partial a'}
(a\cdot a')\otimes (A\wedge A') \;\; ,
\label{product}
\end{equation}
for forms of fixed $Z_{2}$ grading $\partial A$, $\partial A'$ and
matrices of fixed $Z_{2}$ grading $\partial a$ and $\partial a'$.
Linearity allows general products of arbitrary elements to be
constructed. The generalized curvature two-form can now be defined as
${\cal{F}} = dw + w\odot w,$
from which a Lagrangian can be constructed from the norm square via
the scalar product on the algebra of differential forms and the trace
on the space of $2\times 2$  matrices
\begin{equation}
{\cal{L}} = ||{\cal{F}} ||^{2} = Tr <{\overline{\cal{F}}}, {\cal{F}} >
\;\; .
\end{equation}
The result is a Lagrangian consisting of field strength tensors
corresponding to the fields $A$ and $B$ and a Higgs potential in a
symmetry breaking form. This procedure follows also in the case that
$A$ and $B$ are Lie algebra valued, in which case the $2\times 2$
matrix connection is made of $n\times n$ blocks, in general.

An alternate approach, recently investigated by Balakrishna, G\"ursey
and Wali
[2],
which is closer to the original Kaluza-Klein
idea, is to extend space-time by finite matrices.
The extension on the algebra of smooth functions of space-time, $C$,
is an associative but noncommutative algebra
${\cal{A}} = C\otimes M_{n}$,
where $M_{n}$ is the algebra of $n\times n$ matrices, with the
generalized Gell-Mann matrices $\lambda_{a} , a = 1,2,....,n^{2}-1$ as
a hermitian basis. In
analogy to $dx^{\mu}$'s which constitute a basis set on the space of
one-forms on space-time, anticommuting and
anti-hermitian objects, $\Theta_{a}$'s
$a=1,2,....,n^{2}-1$, are introduced. An exterior algebra can then be
defined
\begin{equation}
M_{n}^{*} = M_{n}^{0}\oplus M_{n}^{1}\oplus ......M_{n}^{n^{2}-1} \;\;
,
\end{equation}
where $M_{n}^{p}$ consists of objects of the form
\begin{equation}
F_{a_{1}....a_{p}}\Theta_{a_{1}} ....\Theta_{a_{p}} =
(f_{a_{1}....a_{p} 0}\lambda_{0} + f_{a_{1}....a_{p}
a}\lambda_{a})\Theta_{a_{1}}.... \Theta_{a_{p}} \;\; ,
\end{equation}
with $f_{a_{1}....a_{p}}$ scalar functions on space-time. Writing the
exterior algebra of forms on space-time as $C^{*}$, the exterior
algebra associated with $\cal{A}$ follows from the tensor product
$ {\cal{A}}^{*} = C^{*}\otimes M_{n}^{*} \;\; , $
with the space of generalized $p$ forms appearing as
\begin{equation}
{\cal{A}}^{p} = \sum_{k=0}^{p} C^{k}\otimes M_{n}^{p-k} \;\; ,
\end{equation}
where $C^{k}$ and $M_{n}^{l}$ vanish for $k >4$ and $l>n^{2}-1$
respectively. Higgs fields now appear as space-time scalar
coefficients of the $\Theta_{a}$'s and gauge fields as coefficients of
$dx^{\mu}$'s, as usual, thus introducing Higgs fields at the level of
gauge fields.

A differential operator associated with $M_{n}^{*}$ can be
derived from a consideration of an automorphism on $M_{n}$. A BRST
like operator, $Q$, can then be constructed whose particular form is
dependant on the choice of automorphism. The generalized differential
operator, then, has the form
$D = d + Q ,$
with the connection one-form written as
$w = A + \Phi ,$
$A$ being the gauge one form and $\Phi$ the Higgs field, which has the
form
$\Phi = g\Phi_{a} \Theta_{a},$
where $g$ is the coupling constant. The generalized curvature two-form
then appears as
$F = Dw + w^{2},$
from which a Lagrangian can be constructed by considering the scalar
product on mixed forms
\begin{equation}
{\cal{L}} = {1\over {2g^{2}}}Tr(F_{ij}F^{ij}) \;\; .
\end{equation}
summed over the full range. These components are rescaled so as to
adjust the kinetic terms.
The result is a Yang-Mills Lagrangian with a Higgs potential in a
symmetry breaking form. The choice of automorphism on $M_{n}$ can
allow for greater subtlety in the symmetry breaking pattern.

\section{The Model}
In order to describe an
$SU(2)_{L}\otimes SU(2)_{R}\otimes U(1)$ model we will describe
the generalized connection by a $2\times 2$ matrix consisting of
$2\times 2$ matrix elements,
\begin{equation}
w =
\left (
\begin{array}{cc}
\;\;{\bf{A}}_{L} \;\;\;\;\;\; ig\mu^{-1}\phi \\
ig\mu^{-1}{\overline{\phi}} \;\;\;\; {\bf{A}}_{R} \;\;
\end{array}
\right )
\label{matrix}
\;\; ,
\end{equation}
where ${\bf{A}}_{L}$ and ${\bf{A}}_{R}$ are anti-hermitian,
${\bf{A}}_{LR} = A_{LR} + \Phi_{LR}$
and
$\mu$ is a mass
parameter. So that we will obtain the
algebra $Lie[S(U(2)\otimes U(2)\otimes U(1))]$ we impose
$Str (w)=0= tr{\bf{A}}_{L} -tr\bf{A}_{R}$. In that case the matrix,
obtained by charge conjugating the right hand part,
\begin{eqnarray}
\left (
\begin{array}{ll}
\;\; {\bf{A}}_{L} \;\;\;\;\;\; 0 \;\; \\
\;\;\; 0 \;\;\;\; -{\bf{A}}_{R} \;\;
\end{array}
\right )
\in Lie[S(U(2)\otimes U(2)\otimes U(1))] \;\;  \nonumber \\
\subset Lie[SU(4)] \;\; . \;\;\;\;\;\;\;\;\;\;\;\;\;
\end{eqnarray}
We can therefore decompose the traceless anti-hermitian $4\times 4$
matrix (3.2) with respect to this $SU(4)$ embedding.
Consequently, the generalized connection (\ref{matrix})
can be rewritten
as:
\begin{equation}
w= g
\left (
\begin{array}{ll}
1/2(-iC_{L}+\Phi_{L}-iB/{\sqrt{2}}+\Phi_{0}/{\sqrt{2}})
\;\;\;\;\;\;\;\;\;\;\;\; i\mu^{-1}\phi \;\; \\
\;\;\;\;\;\;\;\;\;\;
\;\;\;\;\;\; i\mu^{-1}\overline{\phi} \;\;\;\;\;\;\;\;
1/2(iC_{R}-\Phi_{R}-iB/{\sqrt{2}}+\Phi_{0}/{\sqrt{2}}) \;\;
\end{array}
\right )
\end{equation}
where $-iC_{L}+\Phi_{L}-iB/{\sqrt{2}}+\Phi_{0}/{\sqrt{2}}$,
say, describes the extended
connection on one copy of extended space-time, with the normalization
of the $SU(4)$ algebra given as $Tr(\Lambda^{\alpha} \Lambda^{\beta})
= 2\delta^{\alpha\beta}$ for $SU(4)$ matrices $\Lambda^{\alpha}$.
In particular
\begin{eqnarray}
A_{L,R}= -igA_{\mu L,R} dx^{\mu} &=& -ig{1\over
2}(B_{\mu}(x)\tau_{0}/{\sqrt{2}} +
C_{\mu L,R a}\tau_{a}) dx^{\mu} \;\; ,
\nonumber \\
\Phi_{L,R} = g\Phi_{L,R a}\Theta_{a} &=& g{1\over
2}(\Phi_{0 a}(x)\tau_{0}/{\sqrt{2}} + \Phi_{L,R a b}
\tau_{b})\Theta_{a} \;\; ,
\label{diffstruc}
\end{eqnarray} where we have the natural choice of automorphism:
\begin{eqnarray}
e^{i\alpha_{b}\tau_{b}} \tau_{a} e^{-i\alpha_{c}\tau_{c}}
&=& \tau_{a} + i [\alpha_{b}\tau_{b} , \tau_{a}] + ......
\nonumber \\
&=& \tau_{a} + d\tau_{a} \,\, ;
\end{eqnarray}
i.e. the inner automorphism. This leads to a derivation $E_{a}$ that
acts on elements of $M_{2}$ [2],
\begin{equation}
E_{a}(F) = {m\over 2} [ \tau_{a} , F] \;\; ,
\end{equation}
where $F\in M_{2}$ and $m$ is a mass scale. Note that a unique
coupling constant has been used, which is consistent with the
construction of a single, generalized, connection. The symmetry
breaking scales will then be separated by the mass scales associated
with each Higgs sector.
In order to proceed we
must construct an exterior derivative consistent with our generalized
structure. On each copy of extended space-time we have the exterior
derivative
$D = d + Q$
where $Q$ is the BRST like operator given by
\begin{equation}
 Q = \Theta_{a} E_{a} - {im\over
2}\epsilon_{abc}\Theta_{a}\Theta_{b} {\partial \big/
\partial\Theta_{c}} \;\; ,
\end{equation}
with $Q^{2}=0$ and $d$ is the normal exterior derivative on space-time.
Note that $dx^{\mu}$'s and $\Theta$'s anticommute ensuring that
$D^{2}=0$. The exterior derivative corresponding to the connection
(\ref{matrix}), $\cal{D}$, will be an operator on $2\times 2$ matrices
with generalized differential form coefficients. For an arbitrary
matrix
we have
\begin{equation}
{\cal {D}}
\left (
\begin{array}{ll}
\;\; A \;\;\; C \;\; \\
\;\; E \;\;\; B \;\;
\end{array}
\right )
 = i[\eta_{\gamma},
\left (
\begin{array}{cc}
\;\; A \;\;\; 0 \;\; \\
\;\; 0 \;\;\; B \;\;
\end{array}
\right )
] + i\{ \eta_{\gamma},
\left (
\begin{array}{cc}
\;\; 0 \;\;\; C \;\; \\
\;\; E \;\;\; 0 \;\;
\end{array}
\right )
\} +
\left (
\begin{array}{rr}
\; DA \;\; -DC \; \\
-DE \;\;\; DB \;
\end{array}
\right )
\;\; ,
\end{equation}
which is a direct generalization of the result of Coquereaux,
Esposito-Farese and Vaillant
[5], eqn
(2.3). The
curvature 2-form is then defined as
${\cal {F}} = {\cal {D}} w + w\odot w ,$
where the product $\odot$ follows from (\ref{product})
\begin{eqnarray}
\left (
\begin{array}{ll}
\;\; A \;\;\; C \;\; \\
\;\; D \;\;\; B \;\;
\end{array}
\right )
\odot
\left (
\begin{array}{rr}
\;\; A' \;\;\; C' \;\; \\
\;\; D' \;\;\; B' \;\;
\end{array}
\right ) \;\;\;\;\;\;\;\;\;\;\;\;\; \nonumber \\ \nonumber \\
=
\left (
\begin{array}{cc}
\;\; A\wedge A' - C\wedge D' \;\;\;\; C\wedge B' - A\wedge C' \;\; \\
\;\; D\wedge A' - B\wedge D' \;\;\;\; B\wedge B' - D\wedge C'
\end{array}
\right ) \;\; ,
\end{eqnarray}
for generalized forms of $Z_{2}$ parity 1. We get
\begin{eqnarray}
{\cal D} w = g
\left (
\begin{array}{ll}
\;\; D({1\over 2}\{ -iC_{L}-iB/{\sqrt{2}}
+\Phi_{L} + \Phi_{0}/{\sqrt{2}}\} ) -\mu^{-1}(e^{-i\gamma}
{\overline{\phi}} + \phi e^{i\gamma})
\;\;\;\;\;\;\;\;\;\;\;\;\;\;\;\;\;\;\;\;\;\;\;\;\;\;\;\;\;\;\;\; \\
\;\; -d(i\mu^{-1}{\overline{\phi}}) +
{1\over 2} (e^{i\gamma}(C_{L}+i\Phi_{L})
+ (C_{R}+i\Phi_{R})e^{i\gamma})
\;\;\;\;\;\;\;\;\;\;\;\;\;\;\;\;\;\;\;\;\;\;\;\;\;\;\;\;\;\;\;\;
\end{array}
\right .
\nonumber \\
\left .
\begin{array}{rr}
\;\;\;\;\;\;\;\;\;\;\;\;\;\;\;\;\;\;\;\;\;\;\;\;\;\;\;\;\;\;
 -d(i\mu^{-1}{\phi}
) +{1\over 2}
((-C_{L} -i\Phi_{L})e^{-i\gamma}+ e^{-i\gamma}(-C_{R} -i\Phi_{R}))
\;\; \\
\;\;\;\;\;\;\;\;\;\;\;\;\;\;\;\;\;\;\;\;\;\;\;\;\;\;\;\;\;\;\;\;
 D({1 \over 2}\{iC_{R}-iB/{\sqrt{2}}-\Phi_{R} +\Phi_{0}/{\sqrt{2}}\} ) -
\mu^{-1}(e^{i\gamma}\phi +
{\overline{\phi}}e^{-i\gamma}) \;\;
\end{array}
\right ) \nonumber \;\;
\end{eqnarray}
and
\begin{eqnarray}
w\odot w = g^{2}
\;\;\;\;\;\;\;\;\;\;\;\;\;\;\;\;\;\;\;\;\;\;\;\;\;\;\;\;\;\;\;\;
\;\;\;\;\;\;\;\;\;\;\;\;\;\;\;\;\;\;\;\;\;\;\;\;\;\;\;\;\;\;\;\;
\;\;\;\;\;\;\;\;\;\;\;\;\;\;\;\;\;\;\;\;\;\;\;\;\;\;\;\;\;\;\;\;
\;\;\;\;\;\;\;\;\;\;\;\;\;\;\;\;\;\;\;\;\;\;\;\;\;
\nonumber \\
\left (
\begin{array}{ll}
\;\;{1\over 4}
 (-iC_{L}+\Phi_{L}-iB/{\sqrt{2}}+\Phi_{0}/{\sqrt{2}})\wedge
(-iC_{L}+\Phi_{L}-iB/{\sqrt{2}}+\Phi_{0}/{\sqrt{2}})-
\mu^{-2} \phi{\overline{\phi}}
\;\;\;\;\;\;\;\;\; \\
\;\; {i\mu^{-1}\over 2}
({\overline{\phi}} [-iC_{L}+\Phi_{L}-iB/{\sqrt{2}}
+\Phi_{0}/{\sqrt{2}}] -
[iC_{R}-\Phi_{R} -iB/{\sqrt{2}}+\Phi_{0}/{\sqrt{2}}]{\overline{\phi}})
\;\;\;\;\;\;\;\;\;
\end{array}
\right .
\nonumber \\
\left .
\begin{array}{rr}
\;\;\;\;\;\;\;\;\;\;\;
{i\mu^{-1}\over 2}
(\phi [iC_{R}-\Phi_{R} -iB/{\sqrt{2}}+\Phi_{0}/{\sqrt{2}}] -
[-iC_{L}+\Phi_{L}-iB/{\sqrt{2}}+\Phi_{0}/{\sqrt{2}}]\phi) \;\; \\
\;\;\;\;\;\;\;\;\;\;\;
{1\over 4}(iC_{R}-\Phi_{R} -iB/{\sqrt{2}}+\Phi_{0}/{\sqrt{2}})
\wedge (iC_{R}-\Phi_{R} -iB/{\sqrt{2}}+\Phi_{0}/{\sqrt{2}})
-\mu^{-2}{\overline{\phi}}\phi \;\;
\end{array}
\right ) \nonumber \;\;
\end{eqnarray}
where $e^{i\gamma}$  belongs to $SU(2)_{L}$ and $SU(2)_{R}$.
This defines the components of the matrix field strength tensor
\begin{equation}
{\cal F} =
\left (
\begin{array}{cc}
\;\; {\cal F}_{11} \;\;\;\; {\cal F}_{12} \;\; \\
\;\; {\cal F}_{21} \;\;\;\; {\cal F}_{22} \;\;
\end{array}
\right ) \;\; .
\end{equation}

We can now construct the Lagrangian by using the scalar product on the
algebra of differential forms and the trace on the space of $2 \times
2$ matrices. This scalar product is on the extended differential
structure described by (\ref{diffstruc}), which is taken to be zero
between forms of different, generalized, order, including zero forms.
Furthermore, the mass parameter $\mu$ is inserted in the
relevant places in order to keep the dimensions correct
[5]. We use $\mu$,
rather than $m$, since $1/\mu$ describes the constant distance
separating our two parallel universes, resulting in a $Z_{2}$,
left$\leftrightarrow$right, symmetry. Note that use
has been made of the abelian nature of $B$ and $\Phi_{0}$
and the shift [2]
$ \Phi_{LR} \rightarrow \Phi_{LR} -(m/2)\tau_{a}\Theta_{a} $,
has been performed in order to remove the nonhomogenous part of the
gauge transformation on each copy of space-time.
The Lagrangian takes the form
\begin{equation}
{\cal{L}} = {1\over 2g^{2}}|| {\cal{F}} ||^{2} =
{1\over 2g^{2}}( || {\cal{F}}_{11} ||^{2} + || {\cal{F}}_{12} ||^{2} +
|| {\cal{F}}_{21} ||^{2} + || {\cal{F}}_{22} ||^{2}) \;\; ,
\end{equation}
where
\begin{eqnarray}
{1\over 2g^{2}}|| {\cal{F}}_{11} ||^{2}
 =-{1\over 8} Tr(F_{\mu\nu L}F^{\mu\nu}_{L}) +
Tr(D_{\mu}(\Phi_{La})D^{\mu}(\Phi_{La}))-V_{L} \nonumber \\
+{1\over 2}Tr(\mu(e^{i\gamma}\phi + {\overline{\phi}} e^{-i\gamma})
 + g\phi
{\overline{\phi}})^{2}, \;\;\;\;\;\;\;\;\;\;\;\;  \nonumber \\
{1\over 2g^{2}}|| {\cal{F}}_{12} ||^{2}
 ={1\over 2}Tr \{ ({\overline{\nabla_{\mu} \phi}}
-{\mu\over 2}(e^{i\gamma}\Phi_{La} + \Phi_{Ra}e^{i\gamma}) -
{g\over 2}({\overline{\phi}}\Phi_{La} +
\Phi_{Ra}{\overline{\phi}})) \;\; , \nonumber \\
 ( {\nabla_{\mu} \phi}
+{\mu\over 2}
(\Phi_{La}e^{-i\gamma} + e^{-i\gamma}\Phi_{Ra}) + {g\over 2}
(\Phi_{La}\phi +
\phi\Phi_{Ra})) \} \nonumber \\
{1\over 2g^{2}}|| {\cal{F}}_{21} ||^{2}
 = {1\over 2}Tr \{ ({\overline{\nabla_{\mu} \phi}}
-{\mu\over 2}(e^{i\gamma}\Phi_{La} + \Phi_{Ra}e^{i\gamma}) -
{g\over 2}({\overline{\phi}}\Phi_{La} +
\Phi_{Ra}{\overline{\phi}})) \;\; , \nonumber \\
 ( {\nabla_{\mu} \phi}
+{\mu\over 2}
(\Phi_{La}e^{-i\gamma} + e^{-i\gamma}\Phi_{Ra}) + {g\over 2}
(\Phi_{La}\phi +
\phi\Phi_{Ra})) \} \nonumber \\
{1\over 2g^{2}}|| {\cal{F}}_{22} ||^{2}
 =-{1\over 8} Tr(F_{\mu\nu R}F^{\mu\nu}_{R}) +
Tr(D_{\mu}(\Phi_{Ra})D^{\mu}(\Phi_{Ra}))-V_{R} \nonumber \\
+{1\over 2}Tr(\mu(e^{i\gamma}\phi + {\overline{\phi}} e^{-i\gamma}) + g
\phi
{\overline{\phi}})^{2}, \;\;\;\;\;\;\;\;
\end{eqnarray}
with
\begin{eqnarray}
\nabla_{\mu}\phi = \partial_{\mu}\phi -
{i\mu\over 2}C_{\mu L} e^{-i\gamma}
-{i\mu\over 2}
e^{-i\gamma} C_{\mu R} -{gi\over 2}
C_{\mu L}\phi -{gi\over 2}\phi C_{\mu R} \;\; ,
\;\;\;\;\;\;\;\;\;\;\;\;\;\;\;\;\;\;\;\;\;\;\;\;\;\;\;
\;\;\;\;\;\;\;\;\; \nonumber \\
\nabla_{\mu} {\overline{\phi}} = {\overline{\nabla_{\mu}\phi}} \;\; ,
\;\;\;\;\;\;\;\;\;\;\;\;\;\;\;\;\;\;\;\;\;\;\;\;
\;\;\;\;\;\;\;\;\;\;\;\;\;\;\;\;\;\;\;\;\;
\;\;\;\;\;\;\;\;\;\;\;\;\;\;\;\;\;\;\;\;\;\; \nonumber \\
-igF_{\mu\nu L} =-ig[
(\partial_{\mu} C_{\nu L} -\partial_{\nu} C_{\mu L} -ig
[ C_{\mu L}, C_{\nu L} ]) +
{1/{\sqrt{2}}}(\partial_{\mu} B_{\nu}
+ \partial_{\nu} B_{\mu})]  ,
\;\;\;\;\;\;\;\;\;\;\;\;\;\;\;\;\;\;\;\; \nonumber \\
-igF_{\mu\nu R} =-ig[
(\partial_{\nu} C_{\mu R} -\partial_{\mu} C_{\nu R} -ig
[ C_{\mu R}, C_{\nu R} ]) +
{1/{\sqrt{2}}}(\partial_{\mu} B_{\nu}
+ \partial_{\nu} B_{\mu})]  ,
\;\;\;\;\;\;\;\;\;\;\;\;\;\;\;\;\;\;\;\; \nonumber \\
gD_{\mu} \Phi_{L a} = g(\partial_{\mu} {\Phi_{a} /{\sqrt{2}}} +
\partial_{\mu}\Phi_{L a} -ig [ C_{\mu L},
\Phi_{L a} ]) \;\; , \;\;\;\;\;\;\;\;\;\;\;\;\;\;\;\;\;
\;\;\;\;\;\;\;\;\;\;\;\;\;\;\;\;\;\;\;\;\;  \nonumber \\
V_{L} = 1/8 Tr \{ ( m\epsilon_{abc}(\Phi_{0c}/{\sqrt{2}} + \Phi_{Lc})
 +ig[\Phi_{L a}, \Phi_{L
b}])(m\epsilon_{abd}(\Phi_{0d}/{\sqrt{2}} + \Phi_{Ld})
 + ig[\Phi_{L a}, \Phi_{L b} ]) \}
\;\;\;\;\;\;\;\;\; \nonumber \\
\end{eqnarray}
where $V_{R}$ and $D_{\mu} \Phi_{R}$ can be obtained by substituting
$-\Phi_{R}$ for $\Phi_{L}$. We can avoid cancellation of
phases in the Lagrangian by writing  $C_{\mu R} e^{-i\gamma}$ and
$\Phi_{Ra} e^{-i\gamma}$.
The generalized connections on each copy of space-time,
${\bf{A}}_{LR}$,
can be rescaled
by introducing a factor of $i$ so that this structure, represented in
(\ref{matrix}), will be hermitian. Furthermore, we could rescale the
field strength components, as done by Balakrishna et al.
[2], so as to adjust the kinetic terms.

Unlike in the $SU(2)\otimes U(1)$ model considered by Coquereaux
et al.[5]
, we are not compelled to have a line and row of zeros in our
$2\times 2$ matrix connection (\ref{matrix}). This arises in their
model from their
choice of form for the Higgs field, $\phi$, connected with the
embedding
$ Lie[S(U(2)\otimes U(1))] \subset Lie[SU(3)] $.
In this regard, our embedding in $SU(4)$ and consideration of a
left$\leftrightarrow$right symmetry has made more efficient use of
this approach. This then implies that the Higgs field $\phi$ will take
the form
\begin{equation}
\phi =
\left (
\begin{array}{cc}
\;\phi_{1}^{0} \;\;\;\; \phi_{1}^{+} \\
\; \phi_{2}^{-} \;\;\;\; \phi_{2}^{0}
\end{array}
\right ) \;\; ,
\end{equation}
so filling the whole matrix connection. Such a bidoublet field
transforms under left and right gauge transformations as
$ \phi \rightarrow U_{L}^{\dagger}\phi U_{R}  ,$
i.e. it is connected to both sectors, an observation which is
consistent with our geometrical approach and the role of the field
$\phi$ in connecting the two space-times. Being in the form of a
generalized standard model Higgs field, this field will be
associated with electroweak symmetry breaking, leaving strong breaking
of the left and/or right sector to $\Phi_{L}$ and/or $\Phi_{R}$. For
our particular choice of automorphism on the algebra $M_{2}$,
$\Phi_{L}$ and $\Phi_{R}$ transform as $SU(2)$ triplets with zero
$U(1)$ charge. Consequently, no mixing occurs at the first stage of
symmetry breaking.
We note that the mass correction can be written as
\begin{equation}
\mu \rightarrow M =
\left (
\begin{array}{cc}
\; \mu \;\;\;\; 0 \; \\
\; 0 \;\;\;\; \mu \;
\end{array}
\right ) \;\; ,
\end{equation}
which is a more natural form than that which appears in the
$SU(2)\otimes U(1)$ approach [5], where it takes the form
\begin{equation}
M =
\left (
\begin{array}{cc}
\;\; \mu \;\;\;\; 0 \; \\
\;\; 0 \;\;\;\; 0 \;
\end{array}
\right )
\end{equation}
which, again, is connected with the choice of form of the Higgs field
$\phi$.

Symmetry breaking patterns for models of the form
$SU(2)_{L}\otimes SU(2)_{R}\otimes U(1)$ have been thoughourly
investigated in the literature
[6]. We note in our model that a
more subtle symmetry breaking scheme can be imposed at the beginning
via the choice of automorphism on $M_{2}$. It is found, in models
incorporating the approach of Balakrishna et.al. [2]
, that those
generators which survive symmetry breaking satisfy
$D(\lambda_{a})=0 \;\; $.
Thus, to ensure mixing between $SU(2)_{R}$ and $U(1)$, say, upon
symmetry breaking, we could follow this approach
and assume the existence of an exterior derivative which satisfies
$D(\tau_{0} + \tau_{3}) = 0 \;\;$.
This then requires a more involved automorphism structure
on $M_{2}$, making use of an inner automorphism on $M_{3}$ [2]. In the
basis of $U$ and $V$ spins this automorphism takes the form
\begin{eqnarray}
e^{i\alpha_{b} U_{b}} \tau_{a} e^{-i\alpha_{c} U_{c}}
&=& \tau_{a} +i[\alpha_{b} U_{b}, \tau_{a} ] + .......
\nonumber \\
&=& \tau_{a} + d\tau_{a}
\end{eqnarray}
where the $\tau_{a}$'s are identified with the $I$ spin matrices. We
can embed this into the $Z_{2}$ graded algebra by making use of the
decomposition
\[ Lie[SU(6)] \supset Lie[S(U(3)\otimes U(3) \otimes U(1))] \;\; , \]
noting that the model is concerned with the algebra $M_{2}$ and not
$M_{3}$  on each copy of space-time. In this respect the approach to
the embedding is different to the previous example, the $U$ and $V$
basis is chosen for convenience only. We identify $\tau_{0}$ with
${1/\sqrt{3}}(\lambda_{8} +
2\Lambda_{0, (6\times 6)})$ ($-\lambda_{8}$
for the right sector), where $\Lambda_{0, (6\times 6)}$ appears in the
decomposition of $Lie[SU(6)]$ as
\begin{equation}
 {1/{\sqrt{3}}}\; diag (1,1,1,-1,-1,-1)
\;\; .
\end{equation}
It is a nice feature that the normalization factors on the diagonal
generators are consistent. The elements of the generalized gauge
connection on each copy of space-time take the form [2]
\begin{eqnarray}
A_{LR} = -ig{1/2} ( B_{\mu} \tau_{0} + C_{\mu LR a} \tau_{a} ) dx^{\mu}
\;\; , \;\;\;\;\;\;\;\;\;\;\;\;\;\;\;  \nonumber \\
H_{LR} = H_{+ LR} V_{+} + H_{0 LR} U_{+} \;\; , \;\; \Delta = {1/2}
(\Delta_{0} \tau_{0} + \Delta_{a LR} \tau_{a} ) \;\; ,
\end{eqnarray}
where $H$, $\Delta_{0}$ and $\Delta_{a}$ constitute the Higgs fields
transforming as a doublet, with $U(1)$ charge 1, and a singlet and
triplet, with $U(1)$ charge zero, under $SU(2)$ respectively.
These quantum numbers follow
from the decomposition of the adjoint of $SU(3)$ under $SU(3) \supset
SU(2)\otimes U(1)$,
\begin{equation}
{\underline{8}} = {\underline{1}}(0) + {\underline{2}}(1) +
{\underline{2}}(-1)+ {\underline{3}}(0) \;\; ,
\end{equation}
where the eigenvalues of the $U(1)$ generator are suitably
normalized. We note that doublet Higgs fields with unit $U(1)$ charge
constitute a candidate set for symmetry breaking in
left$\leftrightarrow$right symmetric models [6,7]. The triplet Higgs
provides for an additional large mass scale to be introduced.
Elements associated with $\tau_{0}$ do not carry $L,R$ indices so as
to maintain $Str(w)=0$.

It appears now that we will no longer fill the matrix connection by
use of a bidoublet Higgs field taking values in the algebra $M_{2}$
. However, this is an artefact of the
choice of automorphism. That this approach does make maximal use of
the gauge connection is more transparent in the previous example. We
see from the expressions (3.13) that the bidoublet Higgs field carries
zero $U(1)$ charge, just as we require.

We can predict a value for the Weinberg angle by noting that we employ
only one coupling constant. The form of the mass eigenstate
corresponding to the photon can be parameterized in terms of
$\theta_{W}$ as
\[ P = \sin{\theta_{W}} (C^{3}_{L} + C^{3}_{R}) +
B(\cos{2\theta_{W}})^{1/2}
\;\; . \]
Ignoring mixing between the left and right sectors, the first stage of
symmetry breaking will result in a $45^{o}$ mixing angle. Making use
of the above parameterization, this implies the result
\begin{equation}
{\sin}^{2} {\theta_{W}} = 1/3 \;\; .
\end{equation}
It is interesting that this falls between the classical
predictions given by Coquereaux et al., $1/4$, [5]
and Balakrishna et al.,
$1/2$ [2]. If we choose the mass scale defined by the triplet Higgs to
correspond to the gauge coupling unification scale we can modify this
result by considering the evolution of the coupling constants to the
electroweak scale. We note that in left$\leftrightarrow$right
symmetric G.U.T. models a scale of $10^{12}$ Gev provides a lower limit
to avoid the domain wall problem. Evolving from this scale, taking an
intermediate left$\leftrightarrow$right symmetric breaking scale of
$10^{6}$ Gev, corresponding to a light $M_{W_{R}}$ model, we get
${\sin}^{2} {\theta_{W}} = 0.23$. We find that evolving from the
Planck scale will not produce a reasonable result. This is not
unexpected as an intermediate mass scale would be expected before the
G.U.T. scale in a true grand unified left$\leftrightarrow$right
symmetric model.

Extending the model of Balakrishna et al. [2] to include fermions via
supersymmetry drastically changes the results in the bosonic
sector [8].
In particular, those
fields responsible for the Higgs potential are excluded from the
Lagrangian. The addition of fermions in the model of Coquereaux et al.
[5] presents less of a challenge. It seems then that an
interesting extension to our model would be an investigation on the
fermionic sector and if it allows for this extension in a natural way.

An interesting point which is raised by this approach is the
homogeneity of forms appearing in the generalized connection
(\ref{connect}) [2].
It has been pointed out
that such matrices define the supergroup $U(n/n)$, for $A$ and
$B$ $n\times n$ matrices
[9]. This allows the differential of two
points to be treated as the differential in a Grassmannian direction.
Taking points of the form $(x_{\mu}, \theta)$ one then writes
$A(x,\delta \theta) = \phi d\theta . $
This opens the possibility that all Higgs fields in our model can be
associated with anticommuting Grassman coordinates, which are not
necessarily identified. In this way the homogeneity of forms in the
connection (\ref{connect}) could be preserved.

\section{CONCLUSION}
We have shown that by allowing the gauge field elements in the
$2\times 2$ matrix connection
to be described by the noncommutative algebra of forms,
that the bosonic sector
of $SU(2)\otimes SU(2)\otimes U(1)$ models can be reproduced which
incorporates an extended Higgs sector necessary to produce parity
violating minima.
Furthermore, the symmetry breaking scheme can be adjusted within this
geometrical approach so as to fine tune the model. Without the
additional Higgs fields provided by the extended algebra of forms on
each copy of space-time, such a left$\leftrightarrow$right symmetric
model could not be considered. By including such fields, full use can
be made of the $2\times 2$ matrix connection allowing the $Z_{2}$
graded
symmetry provided by such a model to be utilized. The form of the
Higgs potential is such that many gauge invariant terms do not appear,
thus reducing the complexity of the most general models.
That Higgs fields may be intimately connected with Grassmannian
coordinates provides a possible avenue for further investigation into
such unified geometrical approaches.

\section{ACKNOWLEDGEMENTS}
The authors would like to thank J. Choi for helpful discussions.
Also, B.E.H would like to acknowledge the support of the Australian
Postgraduate Research Program.


\begin{references}
\bibitem[1]{forg} Forgacs P. \ and  Manton N. \ S. \, {\it
Comm. \ Math. \
Phys.  } {\bf 72}, 15 (1980); Chapline G. \  and  Slansky R. \, {\it
 Nuc. \
Phys. \ B} {\bf 209}, 461 (1982); Manton N. \ S. \, {\it
Nuc. \ Phys. \ B} {\bf
193}, 502 (1981); Chapline G. \  and Manton N. \ S. \, {\it
Nuc. \ Phys. \
B} {\bf 184}, 591 (1981).
\bibitem[2]{balak} Balakrishna B. \ S., \  G\"ursey F. \
and Wali K. \ C., \
 {\it Phys. \ Rev. \ D} {\bf 44}, 3313 (1991); ibid, {\it
 Phys. \ Lett. B} {\bf
254}, 430 (1991). See also Dubois-Violette M., \  Kerner R. \ and
Madore J., {\it Class. and Quant. Grav. }
{\bf 6}, 1709 (1989); ibid,
{\it J. \ Math. \ Phys. } {\bf 31}, 323 (1990).
\bibitem[3]{con} Connes A. \ in {\it The Interface of Mathematics and
Particle Physics}, Edited by D. \ Quillen, G. \ Segal and S. \ Tsou
(Oxford University Press, 1990).
\bibitem[4]{con2} Connes A. \ and Lott J., {\it
Nuc. \ Phys. \ B (proc.
suppl.)} {\bf 18}, 29 (1990).
\bibitem[5]{coq}
 Coquereaux R., \  Esposito-Farese G. \  and Vaillant G. \,
 {\it Nuc. \ Phys. \ B} {\bf 353}, 689 (1991).
\bibitem[6]{hig}See for example Pati J. \ and Salam A., \  {\it
 Phys. \ Rev.
\ D} {\bf 10}, 275 (1974); Mohapatra R. \ and Pati J., \ {\it
Phys. \ Rev. \
D} {\bf 11}, 566 (1975); Senjanovic G. \ and Mohapatra R., \ {\it
 Phys. \ Rev.
\ D} {\bf 12}, 1502 (1975); Mohapatra R. \ and Sidhu D., \ {\it
 Phys. \ Rev. \
Lett. } {\bf 38}, 667 (1977); Mohapatra R. \ and Senjanovic  G., \
{\it
 Phys.
\ Rev. \ Lett. } {\bf 44}, 912 (1980) and
 Gunion J. \ F., \ Grifols J., \
 Mendez A., \ Kayser B. \  and Olness F. \,
 {\it Phys. \ Rev. \ D} {\bf 40},
1546 (1989).
\bibitem[7]{choi}
  Choi J. \ and Volkas  R. \ R., \  {\it Phys. \ Rev. \ D} {\bf 45},
4261 (1992).
\bibitem[8]{balak2}Balakrishna B. \ S., \ G\"ursey F., \ Nguyen Ai Viet
and Wali K. \ C., \ Syracuse University preprint SU-4240-501 (1992).
\bibitem[9]{hus}
Hussain F. \  and Thompson G., \ {\it
Phys. \ Lett. \ B} {\bf 260}, 359 (1991).
\end{references}
\end{document}